\documentclass[twocolumn,aps,prb,superscriptaddress,floatfix]{revtex4}

\begin{document}
\title{Strong correlation, Bloch bundle topology, and spinless Haldane-Hubbard model}

\author{Ilya Ivantsov}
\affiliation{Bogoliubov Laboratory of Theoretical Physics, Joint
Institute for Nuclear Research, Dubna, Russia}
\author{Alvaro Ferraz}
\affiliation{International Institute of Physics - UFRN,
Department of Experimental and Theoretical Physics - UFRN, Natal, Brazil}
\author{ Evgenii Kochetov}
\affiliation{Bogoliubov Laboratory of Theoretical Physics, Joint
Institute for Nuclear Research, Dubna, Russia}

\begin{abstract}

Different realizations of the Hubbard operators in different Hilbert spaces give rise to various microscopic lattice electron models driven by strong correlations. In terms of the Gutzwiller projected electron operators,
the most familiar examples are the conventional $t-J$  model and the BCS-Hubbard model at strong coupling.
We focus instead on the spin-dopon
representation of the Hubbard operators. In this case
the no double occupancy (NDO) constraint can be reexpressed as a Kondo interaction. As an explicit example, the effective low-energy
action is derived in terms of  itinerant spinless fermions (dopons) strongly interacting with localized lattice spins. The spontaneous breaking of the time-reversal symmetry reduces that action to the one describing
a spinless version of the Haldane-Hubbard topological theory. Our consideration suggests that the topologically nontrivial $U(1)$ Bloch bundle associated with this model can in principle be realized
dynamically due to the presence of strong correlations even in the absence of any external local or global fluxes.

\end{abstract}
\maketitle

\section{Introduction}

The key point in constructing lattice electron models that incorporate strong electron correlations, is the local NDO constraint  that reduces the original $4d$ Hilbert space to a $3d$ on-site space.
This constraint is an identity in terms of the Hubbard operators, $X$:
\begin{equation}
X_i^{\uparrow\uparrow}+X_i^{\downarrow\downarrow}+X_i^{00}=1,
\label{1.3}\end{equation}
where $X^{pq}=|p\rangle\langle q|,$ with
$p,q=\sigma, 0$ and $\sigma=\uparrow,\downarrow.$
The Hubbard $X$ operators act in the $3d$ on-site Hilbert space
spanned by the spin up and down states $|\sigma\rangle$ and by the vacuum state, $|0\rangle.$
The $X$ operators provide the faithful (one-to-one, isomorphic) representation of the $u(2|1$) algebra generators \cite{wieg}.
In algebraic language, Eq.(\ref{1.3}) is just the first Casimir operator in the fundamental $u(2|1)$ representation which fixes its dimension. Since the $X$ operators fulfil the complicated (anti)commutation relations of the
$u(2|1)$ (super)algebra, one usually proceeds, in practical calculations, by first attempting  to replace the Hubbard operators by suitable combinations of operators with simplest  commutation relations.
To illustrate this point, consider the homomorphic mapping
$$\phi: X^{\sigma 0}_i \to \tilde c^{\dagger}_{i\sigma}=c^{\dagger}_{i\sigma}(1-n_{i-\sigma}),$$
where $c^{\dagger}_{i\sigma}$ is the creation operator of electron on a site $i$ with spin projection $\sigma$ and
$n_i=\sum_{\sigma}c^{\dagger}_{i\sigma}c_{i\sigma}.$ Although the Gutzwiller projector operators $\tilde c_{\sigma}$ fulfil the commutation relations of the Hubbard ($u(2|1)$) superalgebra, they are not isomorphic to the Hubbard operators.  The $X's$ and $\tilde c's$ operators display different number of degrees of freedom.
As a matter of fact the homomorphic image of the first Casimir operator in these new variables becomes,
$$\phi(X_i^{\uparrow\uparrow}+X_i^{\downarrow\downarrow}+X_i^{00})=\sum_{\sigma}\tilde c^{\dagger}_{i\sigma}\tilde c_{i\sigma}+ \tilde c_{i\uparrow}\tilde c^{\dagger}_{i\uparrow}=
1-n_{i\uparrow}n_{i\downarrow}.$$
It is clear that the projected electron operators establish a faithful
representation of the $u(2|1)$ algebra, or in other words, they become isomorphic to the Hubbard operators, provided the local NDO constraint holds,
\begin{equation}
n_{i\uparrow}n_{i\downarrow}=0.
\label{p1}\end{equation}
Only under this condition one can write
$\tilde c^{\dagger}_{i\sigma}=X^{\sigma 0}_i.$ From now on we identify the $\tilde c_i$ operators with the corresponding Hubbard operators implying that they both act in a $3d$ on-site Hilbert space singled out by Eq.(\ref{p1}).

The focus of this paper is primarily theoretical.
We review both the Hubbard and the generalized Kondo models which are shown to be
equivalent to each other at the infinitely strong coupling but generate quite
different physics out of this extreme regime. We discuss how this phenomenon might
be used to explain the origin of charge-order effects in high temperature
superconductors and the origin of magnetic ordering at very low doping.
We end up by showing that, quite surprisingly, the topologically non-trivial
U(1) Bloch bundle and the associated spinless Haldane-Hubbard model
can be realized dynamically in the strong-coupling regime even if external fluxes are not present
in the physical system.

\section{Preliminaries}

Let us start our analysis with the conventional
Hubbard model that describes lattice electrons hopping in a regular lattice in any dimension  subjected
to a on-site Coulomb repulsion:
\begin{equation}
H_{U} = -t\sum_{ij\sigma}
c_{i\sigma}^{\dagger} c_{j\sigma} + U\sum_i
n_{i\uparrow}n_{i\downarrow},
\label{1.0}\end{equation}
where $t$ is the hopping amplitude between nearest neighbouring (nn) sites., and $U$ is the on-site Coulomb repulsion.
Consider now the $U=+\infty $ limit. Since the on-site operator $n_{i\uparrow}n_{i\downarrow}$ possesses only non-negative eigenvalues,
the infinitely strong $U$-coupling enforces the local NDO constraint, i.e., $n_{i\uparrow}n_{i\downarrow}=0.$ In this limit, it then follows that
\begin{equation}
H_{U= \infty}=
-t\sum_{ij,\sigma}\tilde c^{\dagger}_{i\sigma}\tilde c_{j\sigma}=-t\sum_{ij,\sigma}X^{\sigma 0}_iX^{0\sigma}_j .
\label{1.1}\end{equation}

There are many other Hubbard algebra representations, the most popular being the so-called
oscillatory slave-particle representations. The term "oscillatory" refers to the fact that those representations
are constructed in terms of bilinear boson/fermion combinations. Within a standard slave-boson representation such as
$$ \phi(X_i^{\sigma 0}) = f^{\dagger}_{i\sigma}b_i,$$  $f_{i\sigma}$ stands for a fermion
spinful operator and $b_i$ denotes a boson operator that keeps track of the charge degrees of freedom.
This homomorphic mapping becomes again one-to-one, provided the NDO constraint holds. It takes now the form
$\sum_{\sigma}f^{\dagger}_{i\sigma}f_{\sigma}+b^{\dagger}_ib_i=1$. However in contrast with the representation
(\ref{p1}) this constraint cannot be lifted to the action by adding a new term with a global large coupling as suggested by Eq.(\ref{1.0}).

Among several possible Hubbard algebra representations there is one that is most appropriate
to address the physics close to half-filling.  In that regime, there are localized lattice electrons that carry spin degrees of freedom as well as a small number of conduction fermions -- dopons -- which are essentially vacancies hopping along the lattice.
Exactly the same degrees of freedom are relevant for a lattice Kondo model, which suggests that there is a close relationship  between these two model Hamiltonians in this regime. A starting point is therefore to establish a representation of the Hubbard
operators in terms of these localized and itinerant degrees of freedom. Such a representation was proposed in \cite{wen},
\begin{eqnarray}
\phi(X^{\uparrow 0}_i,X^{\downarrow 0}_i)^T=
\frac{1}{\sqrt{2}}(\frac{1}{2}-2\vec S_i\cdot \vec\tau)\tilde d_i,
\label{1.4}\end{eqnarray}
where $^T$ means the transpose and
$\vec\tau$ is the set of Pauli matrices normalized by the condition $\vec\tau^2=3/4$.
In this framework, the localized electron is represented by the lattice spin $\vec S\in su(2)$
whereas  the mobile doped hole (dopon) is described by the projected hole operator,
$\tilde{d}_{i\sigma}=d_{i\sigma}(1-n^d_{i-\sigma})$.
Here $\tilde d=(\tilde d_{\uparrow},\tilde d_{\downarrow})^{t}$. The local dopon number operator is denoted
as $n^d_{i\sigma}.$
In principle, the "tilde" sign over the dopon
operators can now be dropped, as that would cause an error of order $O(n_d^2)$ which is unimportant
in the underdoped regime. More details can be found in \cite{ifk2}.

The set of operators (\ref{1.4}) becomes identical to the full set of the Hubbard operators,
provided \cite{pfk}
\begin{eqnarray}
\vec{S_i} \cdot
\vec{s_i}+\frac{3}{4}n^d_i=0.
\label{1.5}
\end{eqnarray}
Here $\vec s_i$
stands for a spin dopon operator that interacts with the local spin operator through the Kondo-type coupling. It is important that the operator on the left hand side of Eq. (\ref{1.5}) possesses only non-negative eigenvalues, which implies that the local NDO constraint (\ref{1.5}) can be imposed
by adding to the Hamiltonian the term with {\it global} coupling constant, $\lambda\sum_i(\vec{S_i}\cdot
\vec{s_i}+\frac{3}{4}n^d_i),\,\, \lambda\to +\infty.$

In view of that the hopping of the constrained electrons
takes on a form of the modified Kondo model (MKM),
\begin{eqnarray}
H_{\lambda}\,=\, 2t\sum_{ij\sigma}
{d}_{i\sigma}^{\dagger} {d}_{j\sigma}
+ \lambda
\sum_i(\vec{S_i} \cdot
\vec{s_i}+\frac{3}{4}n^d_i),
\label{1.6}
\end{eqnarray}
where $\lambda/t \to\infty$.
In contrast to the conventional Kondo model, there is now an extra term,
$\sim n^d_i$, that comes from the NDO constraint. It regularizes the ground-state energy and it ensures that the theory remains finite as $\lambda\to +\infty.$
Indeed, let us consider the canonical Kondo model with the AF coupling $K>0$,
\begin{equation}
H_{K}= -t\sum_{ij\sigma}
{d}_{i\sigma}^{\dagger} {d}_{j\sigma}
+ K
\sum_i\vec{S_i} \cdot
\vec{s_i}.
\label{1.7}\end{equation}
In the limit $K\gg t$, it takes the form \cite{sigrist}
\begin{equation}
H_{K}= \frac{t}{2}\sum_{ij\sigma}
\tilde {c}_{i\sigma}^{\dagger} \tilde {c}_{j\sigma}
- \frac{3K}{4}\sum_i(1-\tilde n_i) +{\cal O}(t^2/K),
\label{1.8}\end{equation}
where $\tilde n_i=\sum_{\sigma}\tilde {c}_{i\sigma}^{\dagger}\tilde {c}_{i\sigma}=1-\tilde n^d_i.$
It is clear that the ground-state energy in this limit $\sim -K$.
On the other hand, comparing this with Eq.(\ref{1.6}) gives
\begin{equation}
H_{\lambda= \infty}=
-t\sum_{ij,\sigma}\tilde c^{\dagger}_{i\sigma}\tilde c_{j\sigma}=H_{U=\infty}.
\label{1.9}\end{equation}
We thus see that both the standard  Hubbard model (\ref{1.0}) and the MKM (\ref{1.6}) possess the same strong- coupling limit given by Eq.(\ref{1.9}).

\section{Hubbard physics}

To uncover the physics behind both the standard  Hubbard model (\ref{1.0}) and the MKM (\ref{1.6})
one needs to derive both the leading as well as the next-to-leading terms in  the corresponding effective Hamiltonians.
In this section, we discuss the models of strongly correlated electrons that arise from the conventional Hubbard model at a strong on-site Coulomb repulsion.

\subsection{$t-J$ model}

For the reader's convenience and as a check against some well known results,
we first perform an expansion
of the conventional Hubbard model (\ref{1.0}) in inverse powers of $U$.
To go beyond already well-known results,
we present here a derivation in a general case of the so-called mass-imbalanced Hubbard model.
This model implies that the hopping amplitude $t$ is now taken to be spin-dependent, i.e., $t\to t_{\sigma}.$
We thus write,
\begin{equation}
H_{U}^{\uparrow\downarrow} = H_0 + V,
\label{3.0} \end{equation}
with
\begin{equation}
H_0= U\sum_i
n_{i\uparrow}n_{i\downarrow},
\label{3.1} \end{equation}
\begin{equation}
V=-\sum_{ij\sigma}t_{\sigma}
c_{i\sigma}^{\dagger} c_{j\sigma}.
\label{3.2}\end{equation}
In the large-$U$ limit, we can treat $V$ as a small perturbation.
The ground state of $H_0$ contains no doubly occupied sites and it is spanned by
the physical states given by the set:
$$|O_g\rangle =\{|\uparrow \rangle_i,|\downarrow
\rangle_i,|0\rangle_i\}.$$

Let us now define the operator $P=\prod_i P_i$ that projects into the ground state of $H_0:$
$$P_i=|0\rangle_i\langle 0|_i+|\uparrow\rangle_i\langle \uparrow|_i+|\downarrow\rangle_i\langle \downarrow|_i.$$
Making use of convetional perturbation theory, up to second order in $V$, we can define the effective Hamiltonian:\cite{messiah}
\begin{equation}
H_{eff} = PVP+\sum_{{\phi_n}\ne O_g}\frac{PV|\phi_n\rangle\langle \phi_n|VP}{\epsilon_0-\epsilon_n},
\label{3.3}\end{equation}
where $\epsilon_0=0$ is the ground state energy and $|\phi_n\rangle$ is an eigenstate of $H_0$ with eigenvalue
$\epsilon_n.$
Here we have $Pc^{\dagger}_{\sigma}P=|\sigma\rangle\langle 0|=X^{\sigma 0}=\tilde c^{\dagger}_{\sigma}$ and
$Pc_{\sigma}P=|0\rangle\langle \sigma|=X^{0\sigma}=\tilde c_{\sigma},$ so that
\begin{equation}
PVP= -t\sum_{ij\sigma}
\tilde c_{i\sigma}^{\dagger} \tilde c_{j\sigma}.
\label{3.4}\end{equation}

For simplicity from now on we will be considering solely nearest neighbour sites $i$ and $j$ ignoring the $3$-site interactions. In such a case we get the following eigenstates of $H_0$ with eigenvalue $U:$
$$|\phi_1\rangle = |0\rangle_i|\uparrow,\downarrow\rangle_j$$ and $$|\phi_2\rangle=|0\rangle_j|\uparrow,\downarrow\rangle_i.$$
We then have
$$PV|\phi_1\rangle= |\uparrow\rangle_i|\downarrow\rangle_j-|\downarrow\rangle_i|\uparrow\rangle_j.$$ The state
$PV|\phi_2\rangle$ is given by the same expression with the change $i\leftrightarrow j.$
As a result, we get
$$\sum_{\phi_n\ne O_g}PV|\phi_n\rangle\langle \phi_n| VP=$$
$$ 2(X^{\uparrow\uparrow}_iX^{\downarrow\downarrow}_j-X^{\uparrow\downarrow}_iX^{\downarrow\uparrow}_j-
X^{\downarrow\uparrow}_iX^{\uparrow\downarrow}_j +X^{\downarrow\downarrow}_iX^{\uparrow\uparrow}_j).$$

Equation (\ref{3.3}) then yields the $t-J$ mass-imbalanced model Hamiltonian at arbitrary doping,
\begin{eqnarray}
H_{t-J}^{\uparrow\downarrow}
&=& -t_{\uparrow}\sum_{ij} \tilde{c}_{i\uparrow}^{\dagger}
\tilde{c}_{j\uparrow}-t_{\downarrow}\sum_{ij} \tilde{c}_{i\downarrow}^{\dagger}
\tilde{c}_{j\downarrow}\nonumber\\
&+& \sum_{ij} \frac{2(t_{\uparrow}^2+t_{\downarrow}^2)}{U}(Q^z_i \cdot Q_j^z-\frac{\tilde n_i\tilde n_j}{4})\nonumber\\
&+&\sum_{ij} \frac{4t_{\uparrow}t_{\downarrow}}{U}(Q^x_i \cdot Q_j^x+Q^y_i \cdot Q_j^y)
\label {4.1}\end{eqnarray}
if we neglect three-site interactions.
Here $\vec
Q_i=\sum_{\sigma,\sigma'}\tilde {c}_{i\sigma}^{\dagger}\vec\tau_{\sigma\sigma'}\tilde {c}_{i\sigma'}=
\sum_{\sigma,\sigma'}{c}_{i\sigma}^{\dagger}\vec\tau_{\sigma\sigma'}{c}_{i\sigma'}$
is the local electron spin operator.
At $t_{\uparrow}=t_{\downarrow}=:t$  this model reduces to the standard $SU(2)$ invariant $t-J$ model,
\begin{equation}
H_{t-J}
=-t\sum_{ij\sigma} \tilde{c}_{i\sigma}^{\dagger}
\tilde{c}_{j\sigma}+ J\sum_{ij} (\vec Q_i \cdot \vec Q_j-\frac{\tilde n_i\tilde n_j}{4}),
\label {3.5}\end{equation}
with the spin exchange coupling $J=4t^2/U.$ At $U=\infty$ the double occupancy virtual process is totally prohibited leading to $J=0$.
When $t_{\uparrow}\ne t_{\downarrow}$ the $SU(2)$ and the time reversal symmetries are both broken.
The mass-imbalanced Hubbard model with unequal hopping amplitudes can be implemented in an optical lattice
by loading mixtures of ultracold fermionic atoms with different masses \cite{liu, jotzu}.
At half-filling the model (\ref{4.1}) reduces to the spin-1/2 $XXZ$ model studied
in \cite{liu} by making use of the quantum Monte Carlo method.

\subsection{BCS-Hubbard model}

As a second example, let us consider the BCS-Hubbard model,
\begin{equation}
H_{\Delta- U} = \sum_{ij}\Delta_{ij}
(c_{i\uparrow}^{\dagger} c_{j\downarrow}^{\dagger}- c_{i\downarrow}^{\dagger} c_{j\uparrow}^{\dagger}
+ h.c.)    + U\sum_i
n_{i\uparrow}n_{i\downarrow}.
\label{bcs1}\end{equation}
Here $\Delta_{ij}$ denotes a singlet pair amplitude along the nn bond. Away from the antifferomagnetic (AF) long-range ordered phase, this Hamiltonian describes a quantum liquid of singlet-spin pairs.

In the limit $U=\infty$, the electrons are subjected to the NDO constraint.
At large but finite $U$  we arrive at the effective Hamiltonian ($\Delta/U\ll 1)$,
\begin{eqnarray}
H_{\Delta-J-V}
&=&\sum_{ij}\Delta_{ij}(\tilde c^{\dagger}_{i\uparrow}\tilde c_{j\downarrow}^{\dagger}
-\tilde c^{\dagger}_{i\downarrow}\tilde c_{j\uparrow}^{\dagger} +h.c.)\nonumber\\
&+& J\sum_{ij} (\vec Q_i \cdot \vec Q_j-\frac{\tilde n_i\tilde n_j}{4})+
V\sum_{ij} \tilde n_i\tilde n_j,
\label {bcs2}\end{eqnarray}
where the spin exchange coupling becomes $J=\Delta^2/2U$ and the charge exchange $V=\Delta^2/U=2J.$
It is interesting to note that both spin and charge correlations are driven by a unique
parameter, the electron spin singlet amplitude, $\Delta.$
As is well known high temperature superconductivity (SC) can arise
in a quantum spin liquid \cite{and}, provided such a state can be stabilized either by geometrical frustration or by doping.
A quasi long-range SC order was recently discovered in a doped spin liquid using density-matrix renormalization group studies to treat the $t-J$ model with small hole doping
for long $4$ and $6$  leg cylinders\cite{kivelson}.
This is so far the strongest indication of  SC
that has been found on the square lattice of width $> 4$.

In contrast to the approaches that display  the insulating parent state
as a quantum spin liquid,
the strongly coupled BCS-Hubbard model (\ref{bcs2}) reduces at half filling $(n_i=1)$ to the Heisenberg AF
that exhibits an ordered magnetic ground state as observed in experiment,
$$H_{\Delta-J-V} \to H_J=J\sum_{ij} (\vec S_i \cdot \vec S_j-\frac{1}{4}).$$
The present analytical result fully agrees with what comes out by explicitly computing
the wave-function overlap via exact diagonalization in finite clusters.\cite{park}
At finite doping however the strongly coupled BCS-Hubbard model (\ref{bcs2}) generates both the spin exchange term $\sim J$ as well as the charge exchange interaction $\sim V$.

The derivation of the model (\ref{bcs2}) implies that
the  dynamics  of  the  pairing interaction  arises  from  virtual  states,  whose  energies  correspond  to  the  Mott  gap $\sim U$. In this case one might refer to an unretarded pairing, i.e. a pairing without a glue\cite{14}, which differs from the traditional picture of a retarded  pairing  interaction.  For the cuprate materials, the relative weight of the retarded and nonretarded contributions to the paring interaction still remains an open question.

\section{Kondo physics}

In this section, we focus on the physics  of strongly correlated electrons driven by the Kondo interaction.
In the underdoped cuprates, an enhancement of the nonlocal charge-charge repulsion
indicates that the Kondo physics may play an essential role in the PG phase.
Recent experiments point
toward a possibly dominant role of charge order in describing the physics of high temperature superconductors,
essentially driven by {\it dynamically} generated charge fluctuations (see \cite{italian} and references therein).

\subsection{$t-V$ model}

Let us now consider the  mass-imbalanced MKM:
\begin{equation}
H_{\lambda}^{\uparrow\downarrow} = H_0 + V,
\label{3.7} \end{equation}
with
\begin{equation}
H_0= \lambda
\sum_i(\vec{S_i} \cdot
\vec{s_i}+\frac{3}{4}n^d_i),
\label{3.8}\end{equation}
and
\begin{equation}
V= 2\sum_{ij\sigma}t_{\sigma}
{d}_{i\sigma}^{\dagger} {d}_{j\sigma}
\label{3.9}\end{equation}
In the large-$\lambda$ limit, we treat $V$ as a perturbation to $H_0$.
In the limit $\lambda \to \infty,$
the unperturbed Hamiltonian $H_0$ enforces the NDO constraint to the system of the hopping dopons.

The ground state of the MKM is spanned by the states
$$|O_g\rangle =\{|\Uparrow 0 \rangle_i,|\Downarrow 0
\rangle_i,\frac{|\Uparrow \downarrow\rangle_i-|\Downarrow \uparrow\rangle_i}{\sqrt 2}\},$$
making use of the notation  $|\sigma a\rangle$ with
$\sigma=\Uparrow,\Downarrow$ labeling the spin projection of the
lattice spins and $a=0,\uparrow,\downarrow$ labeling the dopon
states (dopon double occupancy is not allowed).

The projector into the ground state of $H_0$ now reads
$$P_i=|vac\rangle_i\langle vac|_i +|\Uparrow 0\rangle_i\langle \Uparrow 0|_i+ |\Downarrow 0\rangle_i\langle \Downarrow 0|_i,$$ where the vacancy state is the Kondo singlet
$$| vac\rangle_i=\frac{|\Uparrow \downarrow\rangle_i
- |\Downarrow \uparrow\rangle_i}{\sqrt{2}}.$$
We further get
$$Pd_{\uparrow i}P=-\frac{|\Downarrow 0\rangle_i\langle vac|_i}{\sqrt 2},
= -\tilde c^{\dagger}_{\downarrow i}/\sqrt 2.$$
In the same way,
$$Pd_{\downarrow i}P=\frac{\tilde c^{\dagger}_{\uparrow i}}{\sqrt 2},\,\,
Pd^{\dagger}_{\uparrow i}P=-\frac{\tilde c_{\downarrow i}}{\sqrt 2},\,\,
Pd^{\dagger}_{\downarrow i}P=\frac{\tilde c_{\uparrow i}}{\sqrt 2}.$$
As a result,
\begin{equation}
PVP=-t\sum_{ij,\sigma}\tilde c^{\dagger}_{i\sigma}\tilde c_{j\sigma},
\label{p}\end{equation}
which again proves (\ref{1.9}).

At this stage the following remarks are in order. The proposed MKM differs from the conventional Kondo model by the addition of the $\frac{3}{4}n^d_i$ term in Eq. (\ref{3.8}). As emphasized earlier, this extra term makes the theory finite in the infinitely strong coupling limit and this results in a three-fold degeneracy of the ground state of the on-site Hamiltonian in (\ref{3.8}).
In contrast, the canonical on-site
Kondo coupling exhibits a non-degenerate ground state (the on-site  spin-dopon singlet) with the eigenvalue $-3/4K, \,\,K\gg t,$ which makes the theory unstable as $K/t \to \infty.$

To evaluate a small departure from the infinite $\lambda$  limit, one needs to take into account
$8$ (virtual) eigenstates of $H_0$ with eigenvalue $\lambda$ that contribute to Eq.(\ref{3.3}).
These are
$$ |\phi_1\rangle = |\Downarrow 0\rangle_i \frac{|\Uparrow \downarrow\rangle_j+|\Downarrow \uparrow\rangle_j}
{\sqrt 2},$$
$$ |\phi_2\rangle = |\Uparrow 0\rangle_i \frac{|\Uparrow \downarrow\rangle_j+|\Downarrow \uparrow\rangle_j}
{\sqrt 2}$$
$$|\phi_3\rangle =|\Downarrow 0\rangle_i|\Uparrow\uparrow\rangle_j, \quad
 |\phi_4\rangle =|\Uparrow 0\rangle_i|\Downarrow\downarrow\rangle_j.$$
The other $4$ states can be obtained through the change $i\leftrightarrow j$.
Equation (\ref{3.3}) then reduces to
\begin{eqnarray}
H_{t-V}^{\uparrow\downarrow}=
&=& -t_{\uparrow}\sum_{ij} \tilde{c}_{i\uparrow}^{\dagger}
\tilde{c}_{j\uparrow}-t_{\downarrow}\sum_{ij} \tilde{c}_{i\downarrow}^{\dagger}
\tilde{c}_{j\downarrow}\nonumber\\
&+&
\sum_{ij}\frac{t_{\uparrow}^2}{\lambda} (\tilde n_i\tilde n_j+ \tilde n_i\tilde n_{j\uparrow})\nonumber \\
&+&
\sum_{ij}\frac{t_{\downarrow}^2}{\lambda} (\tilde n_i\tilde n_j+ \tilde n_i\tilde n_{j\downarrow}),
\label{4.2}\end{eqnarray}
where $\tilde n_{\sigma}=n_{\sigma}(1-n_{-\sigma}).$
This model exhibits a $Z_2$ global invariance $(\uparrow\leftrightarrow\downarrow)$.
It is a useful model to describe a mixture of strongly interacting ultracold fermionic gases in a regime where
charge fluctuations play a dominant role.

At $t_{\uparrow}=t_{\downarrow}=:t$, it reduces to the $t-V$ model  that exhibits the global $SU(2)$ symmetry,
\begin{equation}
H_{t-V}=-t\sum_{ij,\sigma}X^{\sigma 0}_{i}X^{0\sigma}_j + V\sum_{ij} \tilde n_i\tilde n_j,
\label{3.10}\end{equation}
where $\tilde n_i=\sum_{\sigma}X^{\sigma\sigma}_i.$
The inter-site coupling is given by $V=3t^2/\lambda$ with $V/t\ll 1.$
Here we have dropped an irrelevant constant term $\sim -VN$ where $N$ is a total number of the electrons.
Let us once again notice that in deriving (\ref{4.2}) the three-site terms in the $V$ term have been ignored.
In the context of the Kondo lattice physics, the model (\ref{3.10}) was proposed earlier in \cite{hirsch,lacroix}. We show here that it actually arises as a strong coupling limit
of the {\it modified}  Kondo model (\ref{1.6}) controlled by the NDO constraint. In contrast, in this limit the conventional Kondo model (\ref{1.7}) becomes unstable against $K\to\infty$.
The $t-V$ model  allows for
the following virtual process:
a dopon is transferred from a Kondo spin-dopon singlet at site $i$ to a spin state to form a virtual triplet state, and back to the Kondo singlet at a nearest-neighbour site $j$.
Since a spin-dopon singlet carries charge and no spin this process gives rise to an effective charge-charge exchange
interaction-- a repulsion
between two nearest-neighbour states.
This is in a clear contrast with the $t-J$ model where the opposite spin singly-occupied configurations are connected by virtual excitations to and from a doubly occupied state  to produce the inter-site AF spin exchange coupling.
Within the $t-V$ model, the inter-site repulsion emerges even if a direct Coulomb interaction is ignored.
It is just the manifestation of short-range strong electron correlations which give rise to both spin-spin and charge-charge inter-site couplings.

The Gutzwiller projector operators and the projector operators given by Eq.(\ref{1.4}) are isomorphic to the Hubbard operators and, hence,  to each other provided the NDO constraint is rigorously imposed.
Because of this, the Hubbard model (\ref{1.0}) at $U=\infty$ and the Kondo-lattice model (\ref{1.6}) at $\lambda=\infty$ are isomorphic
to each other. They merge into one and the same limiting form given by Eq.(\ref{1.1}).
However
at finite couplings the NDO constraint is relaxed and the isomorphism between the two sets of operator
no longer holds.
As a result the Hubbard model and the MKM display different physics.
Those models, at finite couplings, are not isomorphic to each other and cannot therefore be mapped onto each other.
The spin-spin exchange $\sim J$ enhances the AF spin correlations, whereas
the inter-site repulsion $\sim V$ that arises in the MKM at strong coupling enhances the formation of charge correlations.
Indeed,
guided by the unusual behavior of the electronic entropy revealed by specific heat measurements, it was recently proposed that
the Kondo physics may play an essential role in the formation of the PG phase in the underdoped cuprates.\cite{cooper}

\section{Effective action and topology}

Direct treatment of the $t-V$ model (\ref{3.10}) written down in terms of the Hubbard operators poses a severe
technical problem due to the complicated  (anti)commutation relations of the underlying $u(2|1)$ superalgebra.
However it is necessary to impose the local NDO constraint rigorously at the very beginning prior to any approximation, e.g. in a mean field treatment, to avoid incorrect physical results.
Because of this we focus on the effective low-energy action of the model in terms of the variables that resolve
the NDO constraint explicitly.

\subsection{Effective action}

In this section, we derive a low-energy effective action of the $t-V$ model (\ref{3.10})
in the $su(2|1)$ coherent-state (CS) basis. Such a representation is formulated in terms of the
variables that explicitly resolve the constraint.
The partition function
takes the form of the
CS phase-space path integral (see Appendix
B):
\begin{equation}
Z_{t-V}=\int D\mu (z,\xi )\ e^{S_{t-V}}, \label{5.1}
\end{equation}
where
$$D\mu (z,\xi )=\prod_{i,t}\frac{d\bar z_i(t)dz_i(t)}{2\pi
i(1+|z_i|^2)^2}\, d\bar\xi_i(t)d\xi_i(t).$$ Here $z_i$ is a
complex number that keeps track of the spin degrees of freedom,
while $\xi_i $ is a complex Grassmann parameter that describes the
charge degrees of freedom. As $\xi_i^2=0$, the NDO constraint is resolved explicitly.
In contrast to the slave-particle
representations, the $u(2|1)$ dynamical variables are gauge independent:
no auxiliary degrees of freedom are introduced.

The effective action
\begin{eqnarray}
S_{t-V}&=&\sum_i\int_0^{\beta}ia_i^{(0)}(t)-\sum_i\int_0^{\beta}\bar\xi_i
\left(\partial_t+ia_i^{(0)}\right)\xi_idt \nonumber\\&-&
\int_0^{\beta}H_{t-V}dt \label{5.2}\end{eqnarray} involves
the $u(1)$-valued connection one-form of the magnetic monopole
bundle (see Appendix A) that can formally be interpreted as a spin
"kinetic" term,
$$ia^{(0)}=-\langle z|\partial_t|z\rangle=\frac{1}{2}\frac{\dot{\bar z}z-\bar z\dot
z}{1+|z|^2},$$ with $|z\rangle$ being the $su(2)$ coherent state.
This term is also frequently referred to as the Berry connection.
The dynamical part of the action takes the form
\begin{eqnarray}
H_{t-V}=-t\sum_{ij}\bar\xi_i\xi_j e^{ia_{ji}}
+H.c. +V\sum_{ij}\bar\xi_i\xi_i\bar\xi_j\xi_j, \label{5.3}
\end{eqnarray}
where $$a_{ij}=-i\log \langle z_i|z_j\rangle,\,\,
\langle z_{i}|z_{j}\rangle =\frac{1+\overline{z}%
_{i}z_{j}}{\sqrt{(1+|z_{j}|^{2})(1+|z_{i}|^{2})}}$$
and  $z_i(t)$ and $\xi_i(t)$ are the dynamical fields.

Under a {\it global} $SU(2)$ rotation,
\begin{eqnarray} z_i\rightarrow
\frac{uz_i+v}{-\overline{v}z_i+\overline{u}},
\label{tr}
\end{eqnarray}  we get
\begin{equation}a_i^{(0)}\to a^{(0)}_i-\partial_t\theta_i,\quad
a_{ij}\rightarrow a_{ij}+\theta_j-\theta_i,
\label{5.4}
\end{equation}
where
\begin{eqnarray}
\theta_i=-i\log \sqrt{\frac{-v\overline{z}_i+u}{-\overline{v}z_i+\overline{%
u}}}, \quad \left(\begin{array}{ll}
u & v \\
-\overline{v} & \overline{u}%
\end{array}
\right) \in \mathrm{SU(2)}.
\end{eqnarray}
Note that the $\theta_i$ is an angular variable,
$$\theta_i=-arg(-v\bar z_i+u).$$

The effective action (\ref{5.2}) is invariant under the
$SU(2)$ rotation given by Eqs. (\ref{5.4})
accompanied by the $U(1)$ transformation of the fermionic field,
\begin{equation}\xi_i\rightarrow
e^{i\theta_i }\xi_i.
\label{5.44}
\end{equation}
A "flux"  through a plaquette $\sum_{plaq}a_{ij}$ generated by the $SU(2)$ transformation
remains invariant under (\ref{5.4}).

For further convenience let us consider separately a real and imaginary parts of $a_{ij}$:
 $$ a_{ji}=\phi_{ji}+i\chi_{ji}.$$
We get
$$ \phi_{ji}= \frac{i}{2}\log\frac{1+\bar z_iz_j}{1+\bar z_jz_i},$$ and
$$ \chi_{ji}=-\frac{1}{2}\log \frac{(1+\bar z_iz_j)(1+\bar z_jz_i)}{(1+|z_i|^2)(1+|z_j|^2)}.$$
It can be checked that the $\phi_{ji}$ potential transforms under (\ref{tr}) in the same way as the $a_{ji}$ does,
i.e.
\begin{equation}
 \phi_{ji}\to \phi_{ji}+\theta_i-\theta_j,
\label{5.444}\end{equation}
whereas, in contrast, $\chi_{ji}$ remains intact.
This transformation appears as a gauge fixing by choosing a specific rotationally covariant frame. The dynamical fluxes do not depend on that choice.

The $\chi_{ji}$ potential results in a renormalization of the electron hopping amplitude. Its fluctuations are gapped and thus determine high energy physics. Low-energy physics is governed by the fluctuations of the phase
variable,
$$\phi_{ji}=-arg(1+ \bar z_iz_j)=arg (1+ \bar z_jz_i),$$ defined modulo $(2\pi)$.
It is clear that $\phi_{ji}=-\phi_{ij}$
The potentials $\phi_i^{(0)}:=ia^{(0)}_i$ and $\phi_{ji}$ formally  remind those gauge fields that define a compact $U(1)$ lattice gauge theory.
This is due to the fact that both theories are formulated as quantum $U(1)$ line bundles. The gauge potentials (local connections)
in both theories transform in the same way under a change in the local trivilization.\cite{nakahara} In our case, such a change is caused by a rotation of the underlying base space -- a canonical transformation of the phase space.

\subsection{Topology}

In the low-energy limit, the Hamiltonian function (\ref{5.3}) takes on the form
\begin{eqnarray}
H_{t-V}=-\sum_{ij}t_{ij}\bar\xi_i\xi_j e^{i\phi_{ji}}
+H.c. +V\sum_{ij}\bar\xi_i\xi_i\bar\xi_j\xi_j, \label{5.5}
\end{eqnarray}
This Hamiltonian describes the interaction between the spin textures and the conduction dopons.
It may display  nontrivial topology -- a nonzero Chern number -- only provided that there are at least two filled bands. This is because the Chern number of a full set of bands is zero -- the corresponding vector bundle is always trivial.
We restrict ourselves to the case of two bands with two opposite Chern numbers.
To this end, let us
consider a bipartite $2d$  lattice $L$ which is a direct sum of two sublattices $A$ and $B$, i.e. $L=A \oplus B$.
We then make the following change of variables on the sublattice $B$,
\begin{equation}
z_i\to-\frac{1}{\bar z_i}, \quad \xi_i\to \xi_ie^{-i\theta_i}
 \quad i\in B. \label{5.6}
\end{equation}
Under this
transformation, $\phi_{ji}\to \phi_{ij} +\theta_j-\theta_i$.
The CS image of the on-site electron spin operator changes sign,
$\vec Q_i\to -\vec Q_i$ (see Appendix B).  The effective action Hamiltonian function then
becomes
\begin{eqnarray}
&&H_{t-V}=-t_1\sum_{<ij>}\bar\xi_i\xi_j e^{i\phi_{ji}}
-t_2\sum_{<ij>\in A}\bar\xi_i\xi_j e^{i\phi_{ji}}\nonumber\\
&-&t_2\sum_{<ij>\in B}\bar\xi_i\xi_j e^{-i\phi_{ji}}
+H.c. +V\sum_{<ij>}\bar\xi_i\xi_i\bar\xi_j\xi_j, \label{5.7}
\end{eqnarray} where
the symbol $<ij>$ stands for the nearest neighbour sites.
The resulting total effective action displays rotation, translation and time-reversal symmetries.
Time-reversal invariance implies that the action is invariant under the change
$$z(t) \to -1/\bar z(\beta-t).$$
It is important that the phase $\phi_{ji}$ enters the hopping terms on the $A$ and $B$ sublattices with  opposite signs. This in turn means that the fluxes through elementary plaquettes in the $A$ and $B$  within a unit cell are opposite  to each other.

The Hall conductance $\sigma_{xy}$ by definition changes sign under the time-reversal transformation. A nonzero value can only occur if time-reversal invariance is broken.
To realize a finite quantum Hall response, it is therefore necessary to break time-reversal symmetry.
In the Haldane model for graphene, this is done by inserting local fluxes which sum up to zero over a unit cell. These fluxes can be described by introducing homogeneous phase factors in the second neighbour hopping amplitude, $t_2\to t_2e^{\pm\phi}.$ In the dynamical model (\ref{5.7}) this can be achieved by enforcing the spontaneous breaking of the time-reversal symmetry:
$<\phi_{ji}>=\phi.$ Here the brackets stands for a ground-state averaging.
This  results in the corresponding mean-field operator Hamiltonian:
\begin{eqnarray}
H_{t-V}&=&-t_2\sum_{<ij>\in A}f^{\dagger}_if_j e^{i\phi}
-t_2\sum_{<ij>\in B}f^{\dagger}_if_j e^{-i\phi}\nonumber\\
&+&H.c. +V\sum_{<ij>}f^{\dagger}_if_if^{\dagger}_jf_j, \label{5.8}
\end{eqnarray} where
the $f_i$ stands for an on-site spinless fermion
operator. Since $H_{t-V}^*\neq H_{t-V}$ the time-reversal symmetry is broken: the resulting Hamiltonian acquires chirality. We may set $t_1=0$ since it does not qualitatively affect topological properties of the model. At $V=0$ this model reduces to the non-interacting spinless Haldane model considered in\cite{cayssol}.
It was shown that the system remains gapless if $\phi=0,\pi$. In contrast, for complex hopping
there is a gap $\sim t_2\sin\phi$  in the single-particle density of states at the Fermi level which is indicative of a Chern insulator. The corresponding Chern numbers are $\pm 1$. On the other hand at nonzero $V$ the model $\label{5.8}$ is referred to as the spinless Haldane -- Hubbard model. It was shown that at small $V$ the system is a topological Chern insulator whereas at sufficiently large $V$  a CDW Mott insulator phase sets in instead.\cite{yi}

A few comments concerning the physical meaning of the angle $\phi$ are in order at this stage.
In the Haldane model, the homogeneous phase factors $t_2\to t_2e^{\pm\phi}$
can be accounted for by introducing external local magnetic dipole moments, ordered ferromagnetically normal to the plane
at the center of each hexagonal cell of the honeycomb lattice.\cite{haldane}
In contrast, within our approach, the spontaneous symmetry breaking is supposed to be at work.
Here we propose an alternative route to show that the Haldane-Hubbard model can at least in principle be  realized dynamically even in the absence of any local or global external fields. All this is effectively driven by
strong electron correlations.

\section{Conclusion}

In conclusion,
we propose an unified approach to address distinct aspects of strongly correlated electron systems in different contexts
based on the different representations of the NDO constraint.
To set the stage we first rederive the spin-imbalanced $t-J$ model Hamiltonian to test our approach against well known results.
To proceed further, we then consider a strong coupling expansion of the BCS-Hubbard model.
In contrast to the approaches that display  the insulating parent state of the high temperature SC
as a quantum spin liquid,
the strongly coupled BCS-Hubbard model is shown to reduce at half filling to the Heisenberg AF
that exhibits an ordered magnetic ground state as observed in experiments in lightly doped cuprates.

Our major concern here is the
derivation of the microscopic $t-V$ model that describes constrained lattice electrons subjected to
the resulting dynamically emergent short-range repulsion.
Such a repulsion arises as a direct manifestation of strong electron correlations.
While the conventional $t-J$ model is related to Hubbard-model physics, the $t-V$ model is shown to be driven by the Kondo physics instead. Only in the limit of infinitely strong couplings both models coincide with each other
but generate quite
different physics out of this extreme regime. We discuss how this phenomenon might
be used to explain the origin of charge-order effects in high temperature
superconductors and the origin of magnetic ordering at very low doping.

The low-energy limit of the $t-V$  model is shown to describe a system of itinerant spinless fermions strongly coupled to localized lattice $su(2)$ spins. Under the assumption that there is a spontaneous breaking of time-reversal symmetry the resulting action reduces to the spinless version of the Haldane-Hubbard topological theory.
Quite surprisingly, the topologically non-trivial
$U(1)$ Bloch bundle and the associated spinless Haldane-Hubbard model
can at least in principle be realized dynamically in the strong-coupling regime even if external fluxes are not present in the physical system.

\section{Appendix}

\subsection{$su(2)$ algebra and coherent states}

Consider the $su(2)$ algebra in the lowest $s=1/2$ representation:
\begin{equation}
[S_z,S_{\pm}]=\pm S_{\pm},\quad [S_{+},S_{-}]=2S_z, \quad \vec
S^2=3/4.\label{1a.1}\end{equation} Acting with the ``lowering``
spin operator $S^{-}$ on the ``highest weight`` state
$|\uparrow\rangle$ we get the normalized $su(2)$ CS parametrized by
a complex number $z$
\begin{equation}
|z\rangle=\frac{1}{\sqrt{1+|z|^2}}\exp( zS^{-})|\uparrow\rangle=
\frac{1}{\sqrt{1+|z|^2}}(|\uparrow\rangle +z|\downarrow\rangle).
\label{1a.2}\end{equation} In the basis spanned by the vectors
$|\uparrow\rangle, \,|\downarrow\rangle$ we have
$S_{+}=|\uparrow\rangle|\langle\downarrow|,
\,S_{-}=|\downarrow\rangle|\langle\uparrow|, \,
S_z=\frac{1}{2}(|\uparrow\rangle|\langle\uparrow|-|\downarrow\rangle|\langle\downarrow|).$
The CS symbols of the $su(2)$ generators are then easily evaluated
to be
\begin{eqnarray}
S_{+}^{cl}:&=&\langle z|S_{+}|z\rangle=\frac{z}{1+|z|^2},\quad
S_{-}^{cl}=\frac{\bar z}{1+|z|^2},\nonumber\\
S_z^{cl}&=&\frac{1}{2}\frac{1-|z|^2}{1+|z|^2},\quad \vec
S_{cl}^2=1/4, \quad (\vec S^2)_{cl}=3/4.\label{1a.3}\end{eqnarray}
In the main text, we drop the lowercase index $"cl"$ whenever this causes no confusion.

There is a one-to-one correspondence between the $su(2)$ generators
~(\ref{1a.1}) and their CS (classical) symbols given by Eqs.
~(\ref{1a.3}). These symbols form the $su(2)$ algebra under the
Poisson (classical) brackets, $[S^{cl}_z,S^{cl}_{\pm}]_{PB}=\pm
S^{cl}_{\pm},$ etc, where
$$[A,B]_{PB}:=i(1+|z|^2)^2(\partial_z A\partial_{\bar z} B-z \bar
z).$$ Given a Hamiltonian $H=H(\vec S)$, the corresponding
imaginary time phase-space action takes on the form,
\begin{equation}
{\cal S}_{su(2)}(\bar z,z)=-\,\int^{\beta}_0\langle z|
\frac{d}{dt}+H|z\rangle dt, \label{1a.4}\end{equation} with the
kinetic term being given by
$$ia=-\langle z|\frac{d}{dt}|z\rangle= \frac{1}{2}\frac{\dot{\bar z}z-\bar z\dot
z}{1+|z|^2}.$$

From the geometrical viewpoint, the $su(2)$ coherent states
$|z\rangle$ can be thought of as sections of the magnetic monopole
bundle $P(S^2, U(1))$, with the U(1) connection one-form, $ia$,
frequently refereed to as the Berry connection. The base space of that
bundle, the two-sphere $S^2=CP^1$, appears as a classical phase-space of
the spin, whereas its covariantly constant sections, $|z\rangle:\,
\,(d+ia)|z\rangle=0$, form a Hilbert space for quantum spin.

\subsection{su$(2|1)$ superalgebra and coherent states}

Acting with the ``lowering`` superspin operators
$X^{\downarrow\uparrow}$ and $X^{\downarrow 0}$ on the ``highest
weight`` state $|\uparrow\rangle$ we get the normalized  su$(2|1)$
coherent state in the $3d$ fundamental representation,
\begin{eqnarray}
|z,\xi\rangle&=&(1+\bar{z}z
+\bar{\xi}\xi)^{-1/2}\exp\left(zX^{\downarrow\uparrow}+\xi
X^{0\uparrow}\right)|\uparrow\rangle \nonumber\\
&=&(1+\bar{z}z +\bar{\xi}\xi)^{-1/2}(|\uparrow\rangle
+z|\downarrow\rangle+\xi |0\rangle), \label{1b.1}
\end{eqnarray} where $z$ is a complex number, and $\xi $ is a complex
Grassmann parameter. The Grassmann parameter appears here due to
the fact that $X^{\downarrow 0}$ is a fermionic operator in
contrast with the operator $X^{\downarrow\uparrow}$. The product
$\xi X^{0\uparrow}$ represents therefore a bosonic quantity as
required.

At $\xi =0$, the su$(2|1)$ CS reduces to the ordinary $su(2)$ CS,
$|z,\xi=0\rangle\equiv |z\rangle$ ~(\ref{1a.2}), parametrized by a
complex coordinate $z \in$ CP$^1$. In contrast, at $z=0$, it
represents a pure fermionic CS.

The CS symbols of the $X$ operators, $X_{cl}=\langle
z,\xi|X|z,\xi\rangle$, are respectively
\begin{eqnarray}
X^{0\downarrow}_{cl}&=&-\frac{z\bar\xi}{1+|z|^2},\quad
X^{\downarrow0}_{cl}=-\frac{\bar
z\xi}{1+|z|^2},\nonumber\\
X^{0\uparrow}_{cl}&=&-\frac{\bar\xi}{1+|z|^2},\quad X^{\uparrow
0}_{cl}=-\frac{
\xi}{1+|z|^2},\nonumber\\
Q^{+}_{cl}=X^{\uparrow\downarrow}_{cl}&=&\frac{z}{1+|z|^2}\left(1-\frac{\bar\xi\xi}{1+|z|^2}\right),\nonumber\\
Q^{-}_{cl}=X^{\downarrow\uparrow}_{cl}&=&\frac{\bar
z}{1+|z|^2}\left(1-\frac{\bar\xi\xi}{1+|z|^2}\right),\nonumber\\
Q^z_{cl}=\frac{1}{2}(X^{\uparrow\uparrow}_{cl}-X^{\downarrow\downarrow}_{cl})&=&
\frac{1}{2}\frac{1-|z|^2}{1+|z|^2}\left(1-\frac{\bar\xi\xi}{1+|z|^2}\right).
\label{1b.2}\end{eqnarray} Given a Hamiltonian as a polinomial
function of the Hubbard operators,$H=H(X)$, the corresponding
imaginary time phase-space action takes on the form,
\begin{equation}
{\cal S}_{su(2|1)}=-\,\int^{\beta}_0\langle z,\xi|
\frac{d}{dt}+H(X)|z,\xi\rangle dt, \label{1b.3}\end{equation} with
the kinetic term given by
\begin{eqnarray}
\langle
z,\xi|(-\frac{d}{dt})|z,\xi\rangle=\frac{1}{2}\frac{\dot{\bar
z}z-\bar z\dot z
+\dot{\bar\xi}\xi-\bar\xi\dot\xi}{1+|z|^2+\bar\xi\xi}.
\label{1b.4}\end{eqnarray}  Substituting  $H(X)$ into
Eq.(\ref{1b.3}) and making the change of variables $z_i\to z_i,\,\,
\, \xi_i\to\xi_i\sqrt{1+|z_i|^2}$, we are led to the effective
action (\ref{5.1}-\ref{5.3}).


\begin{thebibliography}{99}


\bibitem{wieg} P.B. Wiegmann, Phys. Rev. Lett. {\bf 60}, 821 (1988).
\bibitem{wen} T.C. Ribeiro, and X.-G. Wen, Phys. Rev. Lett. {\bf 95}, 057001 (2005);
T.C. Ribeiro, and X.-G. Wen, Phys. Rev. B {\bf 74}, 155113 (2006).
\bibitem{ifk2} I. Ivantsov, A. Ferraz, and E. Kochetov,
Phys. Rev. B {\bf 95}, 155115 (2017).
\bibitem{pfk} A. Ferraz, E. Kochetov, and B. Uchoa, Phys. Rev. Lett. {\bf 98},
069701 (2007); R.T. Pepino, A. Ferraz, and E. Kochetov, Phys.
Rev. B {\bf 77}, 035130 (2008).
\bibitem{sigrist} M. Sigrist, H. Tsunetsugu, K. Ueda, and T.M. Rice, Phys. Rev. B {\bf 46}, 13838 (1992).
\bibitem{messiah} A. Messiah, {\it Quantum Mechanics} (Dover, Mineola, NY, 1999).
\bibitem{jotzu} G. Jotzu, M. Messer, F. G\"org, D. Greif, R. Desbuquois, and T. Esslinger,
Phys. Rev. Lett. {\bf 115}, 073002 (2015).
\bibitem{liu} Ye-Hua Liu and Lei Wang, Phys. Rev. B {\bf 92}, 235129 (2015); and references therein.
\bibitem{and} P.W. Anderson, Science {\bf 235}, 1196 (1987).
\bibitem{kivelson} Hong-Chen Jiang and Steven A. Kivelson, arXiv:2104.01485v1;
arXiv:2105.07048v1.
\bibitem{park} K. Park, Phys. Rev. Lett. {\bf 95}, 027001 (2005).
\bibitem{14}  T.A. Maier, D. Poilblanck, and D.J. Scalapino, Phys. Rev. Lett. {\bf 100}, 237001 (2008).
\bibitem {italian} R. Arpaia and G. Ghiringhelli, arXiv:2106.00731v1.
\bibitem{hirsch} J.E. Hirsch, Phys. Rev. B {\bf 30}, 5383 (1984).
\bibitem{lacroix} C. Bastide and C. Lacroix, Europhys. Lett. {\bf 4}, 935 (1987).
\bibitem{cooper} J.R. Cooper, arXiv:2108.02108v1.
\bibitem{nakahara} M. Nakahara, {\it Geometry, Topology and Physics}, Second Edition, 2 ed., Taylor and Francis, (2003).
\bibitem{cayssol} J. Cayssol, arXiv:1310.0792v1.
\bibitem{yi} T.-C. Yi, S. Hu, E.V. Castro, and R. Mondaini, arXiv:2107.10056v1.
\bibitem{haldane} F.D.M. Haldane, Phys. Rev. Lett. {\bf 61}, 2015 (1988).
\end{thebibliography}
\end{document}